\documentclass[10pt]{article}
\usepackage[dvips]{graphicx}

\usepackage{epsfig}
\usepackage{graphicx}
\usepackage{indentfirst}
\usepackage{amsmath}
\usepackage{amsfonts}
\usepackage{amssymb}

\begin{document}

%%%%%%%%%%%%%%%%%%%%%%%%%%%%%%%%%%%%%%%%%%%%%%%%%%%%%%%%%%%%%%%%%%%%%%%%
\begin{center}
\begin{flushright}\begin{small}    UFES 2011
\end{small} \end{flushright} \vspace{1.5cm}
\huge{New Static Solutions in $f(T)$ Theory} 
\end{center}

\begin{center}
{\small  M. Hamani Daouda $^{(a)}$}\footnote{E-mail address:
daoudah8@yahoo.fr}\ ,
{\small  Manuel E. Rodrigues $^{(a)}$}\footnote{E-mail
address: esialg@gmail.com}\ and
{\small    M. J. S. Houndjo $^{(b)}$}\footnote{E-mail address:
sthoundjo@yahoo.fr} \vskip 4mm

(a) \ Universidade Federal do Esp\'{\i}rito Santo \\
Centro de Ci\^{e}ncias
Exatas - Departamento de F\'{\i}sica\\
Av. Fernando Ferrari s/n - Campus de Goiabeiras\\ CEP29075-910 -
Vit\'{o}ria/ES, Brazil \\
(b) \ Instituto de F\'{i}sica, Universidade Federal da Bahia, 40210-340, Salvador, BA, Brazil\\
\vskip 2mm

\end{center}

%%%%%%%%%%%%%%%%%%%%%%%%%%%%%%%%%%%%%%%%%%%%%%%%%%%%%%%%%%%%%%%%%%%%%%%%%%%%%%%%%%%%%%%%%%%%%%%%%%%%%%%%%%%%%%%%%%%%%%%%%%%%%%%%%%%%%%%%%%%%%%%%%%%%%%%
\begin{abstract}
We consider the equations of motion of an anisotropic space-time in $f(T)$ theory, where  $T$ is the torsion. New spherically symmetric solutions of black holes and wormholes are  obtained with a constant torsion and the cases for which the radial pressure is proportional to a real constant, to some algebraic functions $f(T)$ and their derivatives $f_T(T)$, or vanish identically.

\end{abstract}
%%%%%%%%%%%%%%%%%%%%%%%%%%%%%%%%%%%%%%%%%%%%%%%%%%%%%%%%%%%%%%%%%%%%%%%%%%%%%%%%%%%%%%%%%%%%%%%%%%%%%%%%%%%%%%%%%%%%%%%%%%%%%%%%%%%%%%%%%%%%%%%%%%%%%%%
Pacs numbers: 04.50. Kd, 04.70.Bw, 04.20. Jb
%%%%%%%%%%%%%%%%%

\section{Introduction}

%%%%%%%%%%%%%%%%%%%%%%%%%%%%%%%%%%%%%%%%%%%%%%%%%%%%%%%%%%%%%%%%%%%%%%%%%%%%%%%%%%%%%%%%%%%

Through a Einstein's proposal for finding  a new version  to General Relativity (GR) \cite {case},  an alternative gravitation theory,  namely Teleparallel Theory (TT) has been embraced and abandoned  later for many years. However, from the considerations made by Moller, the proposal for an analogy between  TT and GR has been undertaken and developed once again \cite{moller}. The RG is a theory that describes the gravitation through the space-time curvature. Since a manifold may possess curvature and torsion, as Cartan spaces, one can separate all the  terms coming from the torsion in the  geometrical object as Riemann tensor, connection, etc...  Hence, we mention that the theory that describes the gravitation as the action of space-time curvature, ie,  coming from the Riemann tensor without torsion or without antisymmetric connection, can be similarly viewed as a theory that possesses only torsion and whose Riemann tensor without torsion vanishes identically. \par
With the recent observational data about the evolution  and the content of the universe, as the accelerated expansion, the existence of dark energy and dark matter, several new  proposals for modifying the GR  are being tested. Since the theories of unifications in low-energy scales have in their effective actions terms as  $ R ^ 2 $, $ R ^ {\mu \nu} R_ {\mu \nu} $ and $ R ^ {\mu \nu \alpha \beta} R_ {\mu \nu \alpha \beta} $, a candidate to the modification of the GR that agrees with  the cosmological and astrophysical observational data is the $f(R)$  modified  gravity \cite{capozziello,odintsov,manuel5}. The main problem which appears with the $f(R)$ theory is that the equations of motion are of the order 4, becoming more difficult  to be analysed than the GR. Since the GR has an analogous Teleparallel Theory, it has been thought to use a theory called $f(T)$, with $T$ being the  torsion scalar, which would be the analogous of the generalization of  the GR, namely, the $f(R)$ theory. The $f(T)$ theory is a generalization of the teleparallel one as we shall see later and  also does not possess curvature, nor Riemann tensor coming from the terms without torsion.\par
In cosmology, the $f(T)$ theory was used originally as a source driving the inflation \cite{fiorini}.  Thereafter, it has been used as an alternative proposal for the acceleration of the universe, without requiring the introduction of dark energy  \cite{ferraro,ratbay,eric,bamba,dutta}. In gravitation, the use of  $f(T)$ theory started providing  solutions for BTZ black holes \cite{fiorini2}. Also, using $f(T)$ it is shown that the first thermodynamic law for black hole can be violated due to the lack of local Lorentz invariance \cite{rong}.
Recently, spherically symmetric static solutions are searched in some $f(T)$ models of gravity theory with a Maxwell term \cite{wang}, and the existence of relativistic stars in f(T) modified gravity is examined, constructing explicitly several classes of static perfect fluid solutions \cite{boehmer}.\par
In this paper, as well as in others methods of simplification of the equations of motion used in GR, for getting  solutions with anisotropic symmetry  as in \cite{bayin}, we propose to analyse  the possibilities of obtaining new solutions in gravitation with $f(T)$ theory. \par
The paper is organized as follows.  In the section 2,  we will present a brief revision  of the fundamental concepts of Weitzenbock's geometry,  the action of $f(T)$ theory and the equations of motions. In the section 3, we will fix the symmetries of the geometry and present the equations for the density, the radial and tangential pressures. New solutions  of $f(T)$ theory are obtained in the section 4 for the cases in which the torsion is constant and the radial pressure vanishes. Finally, in the section 5,  our conclusion and perspectives are presented.

%%%%%%%%%%%%%%%%%%%%%%%%%%%%%%%%%%%%%%%%%%%%%%%%%%%%%%%%%%%%%%%%%%%%%%%%%%%%%%%%%%%%%%%%%%%%%%%%%%%%%%%%%%%%%%%%%%%%%%%%%%%%%%%%%%%%%%%%%%%%%%%%%%%%

\section{\large The basic notions and the field equations}

%%%%%%%%%%%%%%%%%%%%%%%%%%%%%%%%%%%%%%%%%%%%%%%%%%%%%%%%%%%%%%%%%%%%%%%%%%%%%%%%%%%%%%%%%%%%%%%%%%%%%%%%%
The Teleparallel Theory appeared to be equivalent to the GR \cite{wanas,nashed,pereira}, and so, we will introduce the basic concepts of the $f(T)$ theory. First, to avoid any confusion, let us define the notation of the Latin subscript as those  related to the tetrad fields, and  the Greek one related to the space-time coordinates. The line element of the manifold  is given by
\begin{equation}
dS^{2}=g_{\mu\nu}dx^{\mu}dx^{\nu}\; .
\end{equation} 
This line element can be converted to a Minkowskian description by the matrix transformation  called tetrad, as follows
\begin{eqnarray}
dS^{2} &=&g_{\mu\nu}dx^{\mu}dx^{\nu}=\eta_{ij}\theta^{i}\theta^{j}\label{1}\; ,\\
dx^{\mu}& =&e_{i}^{\;\;\mu}\theta^{i}\; , \; \theta^{i}=e^{i}_{\;\;\mu}dx^{\mu}\label{2}\; ,
\end{eqnarray} 
where $\eta_{ij}=diag[1,-1,-1,-1]$ and $e_{i}^{\;\;\mu}e^{i}_{\;\;\nu}=\delta^{\mu}_{\nu}$ or  $e_{i}^{\;\;\mu}e^{j}_{\;\;\mu}=\delta^{j}_{i}$. The root of the metric determinant is given by  $\sqrt{-g}=\det{\left[e^{i}_{\;\;\mu}\right]}=e$. 
For a manifold in which the Riemann tensor part without the torsion terms is null (contributions of the Levi-Civita connection) and only the non zero torsion terms exist, the Weitzenbock's connection components are defined as
\begin{eqnarray}
\Gamma^{\alpha}_{\mu\nu}=e_{i}^{\;\;\alpha}\partial_{\nu}e^{i}_{\;\;\mu}=-e^{i}_{\;\;\mu}\partial_{\nu}e_{i}^{\;\;\alpha}\label{co}\; .
\end{eqnarray}
Defining the components of the torsion and the contorsion  respectively as
\begin{eqnarray}
T^{\alpha}_{\;\;\mu\nu}&=&\Gamma^{\alpha}_{\nu\mu}-\Gamma^{\alpha}_{\mu\nu}=e_{i}^{\;\;\alpha}\left(\partial_{\mu} e^{i}_{\;\;\nu}-\partial_{\nu} e^{i}_{\;\;\mu}\right)\label{tor}\;,\\
K^{\mu\nu}_{\;\;\;\;\alpha}&=&-\frac{1}{2}\left(T^{\mu\nu}_{\;\;\;\;\alpha}-T^{\nu\mu}_{\;\;\;\;\alpha}-T_{\alpha}^{\;\;\mu\nu}\right)\label{cont}\; ,
\end{eqnarray}
and the components of the tensor $S_{\alpha}^{\;\;\mu\nu}$ as
\begin{eqnarray}
S_{\alpha}^{\;\;\mu\nu}=\frac{1}{2}\left( K_{\;\;\;\;\alpha}^{\mu\nu}+\delta^{\mu}_{\alpha}T^{\beta\nu}_{\;\;\;\;\beta}-\delta^{\nu}_{\alpha}T^{\beta\mu}_{\;\;\;\;\beta}\right)\;,
\end{eqnarray}
one can write the torsion scalar  as 
\begin{eqnarray}
T=T^{\alpha}_{\;\;\mu\nu}S^{\;\;\mu\nu}_{\alpha}\label{tore}\; .
\end{eqnarray}
Now, similarly to the $f(R)$ theory, one defines the action of $f(T)$ theory being 
\begin{eqnarray}
S[e^{i}_{\mu},\Phi_{A}]=\int\; d^{4}x\;e\left[\frac{1}{16\pi}f(T)+\mathcal{L}_{Matter}\left(\Phi_{A}\right)\right]\label{action}\; ,
\end{eqnarray}
where we used the units in which $G=c=1$ and the $\Phi_{A}$ are matter fields. Considering the action (\ref{action}) as a functional of the fields  $e^{i}_{\mu}$ and $\Phi_{A}$,  and vanishing the variation of the functional  with respect to the field  $e^{i}_{\nu}$, one obtains the following equation of motion  \cite{barrow}
\begin{eqnarray}
S^{\;\;\nu\rho}_{\mu}\partial_{\rho}Tf_{TT}+\left[e^{-1}e^{i}_{\mu}\partial_{\rho}\left(ee^{\;\;\alpha}_{i}S^{\;\;\nu\rho}_{\alpha}\right)+T^{\alpha}_{\;\;\lambda\mu}S^{\;\;\nu\lambda}_{\alpha}\right]f_{T}+\frac{1}{4}\delta^{\nu}_{\mu}f=4\pi\mathcal{T}^{\nu}_{\mu}\label{em}\; ,
\end{eqnarray}
where $\mathcal{T}^{\nu}_{\mu}$ is the energy momentum tensor. Let us consider that the matter content  is an anisotropic fluid, such that the energy-momentum tensor is given by
\begin{eqnarray}
\mathcal{T}^{\,\nu}_{\mu}=\left(\rho+p_t\right)u_{\mu}u^{\nu}-p_t \delta^{\nu}_{\mu}+\left(p_r-p_t\right)v_{\mu}v^{\nu}\label{tme}\; ,
\end{eqnarray}
where $u^{\mu}$ is the four-velocity, $v^{\mu}$ the unit space-like vector in the radial direction, $\rho$ the energy density, $p_r$ the pressure in the direction of $v^{\mu}$ (normal pressure) and $p_t$  the pressure orthogonal to $v_\mu$ (transversal pressure). Since we are assuming an anisotropic spherically symmetric matter, on has $p_r \neq p_t$, such that their equality corresponds to an isotropic fluid sphere.\par
In the next section, we will make some considerations for the manifold symmetries in order to obtain simplifications in the equation of motion and the specific solutions of these symmetries.
%%%%%%%%%%%%%%%%%%%%%%%%%%%%%%%%%%%%%%%%%%%%%%%%%%%%%%%%%%%%%%%%%%%%%%%%%%%%%%%%%%%%%%%%%%%%%%%%%%%%%%%%

\section{\large   Specifying the geometry}

%%%%%%%%%%%%%%%%%%%%%%%%%%%%%%%%%%%%%%%%%%%%%%%%%%%%%%%%%%%%%%%%%%%%%%%%%%%%%%%%%%%%%%%%%%%%%%%%%%%%%%%%%

Assuming that the manifold possesses a stationary  and spherical symmetry, the metric can be written as 
\begin{equation}
dS^{2}=e^{a(r)}dt^{2}-e^{b(r)}dr^{2}-r^{2}\left(d\theta^{2}+\sin^{2}\left(\theta\right)d\phi^{2}\right)\label{ele}\; .
\end{equation} 
In order to re-write the line element (\ref{ele}) into the invariant form under the Lorentz transformations as in  (\ref{1}), we define the tetrad matrix (\ref{2}) as
\begin{eqnarray}
\left[e^{i}_{\;\;\mu}\right]= diag \left[e^{a(r)/2},e^{b(r)/2},r,r\sin\left(\theta\right)\right]\label{tetra}\; .
\end{eqnarray}

Using  (\ref{tetra}), one can obtain $e=\det{\left[e^{i}_{\;\;\mu}\right]}=e^{(a+b)/2}r^2 \sin\left(\theta\right)$, and with (\ref{co})-(\ref{tore}), we determine the  torsion scalar and its derivatives in terms of  $r$ as
\begin{eqnarray}
T(r) &=& \frac{2e^{-b}}{r}\left(a^{\prime}+\frac{1}{r}\right)\label{te}\; ,\\
T^{\prime} (r)&=& \frac{2e^{-b}}{r}\left(a^{\prime\prime}-\frac{1}{r^2}\right)-T\left(b^{\prime}+\frac{1}{r}\right)\label{dte}\; ,
\end{eqnarray} 
where the prime  ($^{\prime}$) denotes the derivative with respect to  the radial coordinate $r$. One can now re-write the equations of motion (\ref{em}) for an anisotropic fluid as 
\begin{eqnarray}
4\pi\rho &=& \frac{f}{4}-\left( T-\frac{1}{r^2}-\frac{e^{-b}}{r}(a'+b')\right)\frac{f_T}{2}\,,\label{dens} \\
4\pi p_{r} &=& \left(T-\frac{1}{r^2}\right)\frac{f_T}{2}-\frac{f}{4}\label{presr} \\
4\pi p_{t} &=& \left[\frac{T}{2}+e^{-b}\left(\frac{a''}{2}+\left(\frac{a'}{4}+\frac{1}{2r}\right) (a'-b')\right)\right]\frac{f_T}{2}-\frac{f}{4}\,, \label{prest}\\
\frac{\cot\theta}{2r^2}T^{\prime}f_{TT}&=&0\,\,\,,\label{impos}
\end{eqnarray} 
where $p_{r}$ and $p_{t}$  are the radial and tangential pressures respectively. In the Eqs. (\ref{dens})-(\ref{prest}), we used the imposition  (\ref{impos}), which arises from the non-diagonal components $\theta-r$ ($2-1$) of the equation of motion (\ref{em}). This imposition does not appear in the static case of the GR, but making its use in (\ref{impos}), we get only the following possible solutions 
\begin{eqnarray}
T^{\prime}&=&0\Rightarrow T=T_0\label{impt1}\;,\\
f_{TT}&=&0\Rightarrow f(T)=a_{0}+a_{1}T\label{impf1}\;,\\
T^{\prime}&=&0,f_{TT}=0\Rightarrow T=T_0,f(T)=f(T_{0})\label{imptf1}\;,
\end{eqnarray}
which always relapse into the particular case of Teleparallel theory, with $f(T)$ a constant or a linear function.\par
In the next section, we will determine new solutions for the $f(T)$ theory making some consideration about the matter components $\rho (r),\, p_{r} (r)$ and $p_{t} (r)$.

%%%%%%%%%%%%%%%%%%%%%%%%%%%%%%%%%%%%%%%%%%%%%%%%%%%%%%%%%%%%%%%%%%%%%%%%%%%%%%%%%%%%%%%%%%%%%%%%%%

\section{Obtaining new solutions}
Several works have been done in cosmology, modeling and solving some problems, using  the f(T) theory as basis.  Actually, in local and astrophysical phenomena, their is still slowly moving to obtain new solutions. Recently, Deliduman and Yapiskan \cite{yapiskan} shown that it could not exist relativistic stars, such as that of neutrons and others, in $f (T)$ theory, except in the linear trivial case, the usual Teleparallel Theory. However, Boehmer et al \cite{boehmer} showed that for the cases where $T=0$ and $ T^{\prime} = 0 $, there exists solutions of relativistic stars. In the same way, we would like to show some classes of spherically symmetric static solutions of the theory coming from $f(T)$.
\begin{enumerate}
%%%%%%%%%%%%%%%%%%%%%%%

\item  Let us start with the simple case, in which the torsion is constant $T=T_{0}$, which satisfies the imposition (\ref{impt1}). In this case, considering the condition 
\begin{equation}
a^{\prime\prime}(r)=\frac{1}{r^2}\label{cond1},
\end{equation}
which generalizes the case of   Boehmer et al \cite{boehmer}, where $T=0$, for
\begin{equation}
a^{\prime}(r)=-\frac{1}{r}+c_{0}\; ,
\end{equation}
one gets 
\begin{equation}
e^{a(r)}=\left(\frac{r_{0}}{r}\right)e^{c_{0}r}\label{a1}\; ,
\end{equation}
where $c_{0}$ is a real integration constant. From (\ref{a1}) and (\ref{te}),  one obtains
\begin{equation}\label{b1}
e^{b(r)}=\left(\frac{2c_{0}}{T_{0}}\right)\frac{1}{r}\; .
\end{equation}
It is important to note that from the equation (\ref{te}) we can distinguish two interesting situations about $T_{0}$. Hence, $c_0 >0$ implies that $T_{0}>0$, while $c_0<0$ leads to $T_0<0$. So, $e^b$ is always positive in (\ref{b1}).\par
The line element (\ref{ele}) is given by
\begin{equation}
dS^{2}=\left(\frac{r_{0}}{r}\right)e^{c_{0}r}dt^{2}-\left(\frac{2c_{0}}{T_{0}}\right)\frac{dr^{2}}{r}-r^2 d\Omega^2\label{sol1}\; .
\end{equation}
Here, the condition $r_{0}>0$ determines the signature of the metric as $diag(+---)$. The energy density, the radial and tangential pressures can be obtained from the expressions (\ref{dens})-(\ref{prest}) as 
\begin{eqnarray}
\rho (r) &=& \frac{f(T_0)}{16\pi}-\frac{f_{T}(T_0)T_{0}}{16\pi}-\frac{f_{T}(T_0)T_{0}}{8\pi c_{0}r}+\frac{f_{T}(T_0)}{8\pi r^2}\label{dens1}\; ,\\
p_{r} (r) &=& \frac{f_{T}(T_0)T_{0}}{8\pi}-\frac{f(T_{0})}{16\pi}-\frac{f_{T}(T_{0})}{8\pi r^2}\label{presr1}\; ,\\
p_{t} (r) &=& \frac{5c_{0}f_{T}(T_0)T_{0}}{64\pi}-\frac{f(T_0)}{16\pi}+\frac{f_{T}(T_0)T_{0}}{32\pi c_{0}r}+\frac{c_{0}f_{T}(T_0)T_{0}r}{64\pi}\label{prest1}\; .
\end{eqnarray} 

It is easy to observe through (\ref{dens1})-(\ref{prest1}) that $\rho(r)$, $p_r(r)$ and $p_t(r)$ diverge as $r$ goes to zero, which means that there is a singularity at the origin. It appears here that the imposition (\ref{impt1}) makes free the algebraic function $f(T)$. Here, we confirm our position against that of Deliduman and Yapiskan \cite{yapiskan}, where there exists solutions of stars for $f(T)$ theory, and (\ref{sol1}) is one of them. This solution can only be analyzed in parts, for example, the exponential term in $g_{00}$ ( term of type Yukawa potential) was obtained in \cite{stelle} and related to the $f(R)$ theory in \cite{mariafelicia}.\par
On the other hand, we can choose another condition for the line element (\ref{ele}). Then, embracing the condition of quasi-global coordinate
\begin{equation}
a^{\prime}(r)=-b^{\prime}(r)\label{cond2}\; ,
\end{equation}
from (\ref{te}), we get the differential equation
\begin{equation}
\left(e^{-b}\right)^{\prime}+\left(\frac{1}{r}\right)\left(e^{-b}\right)-\frac{T_{0}}{2}r=0\label{eq1}\; ,
\end{equation} 
whose the solution yields the line element
\begin{equation}\label{sol2}
dS^{2}=\left(\frac{c_{0}}{r}+\frac{T_{0}}{6}r^2\right)dt^2-\left(\frac{c_{0}}{r}+\frac{T_{0}}{6}r^2\right)^{-1}dr^2-r^2d\Omega^2\; .
\end{equation}
This is similar to Schwarzschild-(Anti)de Sitter's solution when we fix $c_{0}=-2M$ and $T_{0}=-2\Lambda$. Wang \cite{wang} obtained similar solution, in which the unique difference is the fixation of the function $f(T)=T=-2\Lambda$. The energy density, the radial and tangential pressures can be obtained from the expressions (\ref{dens})-(\ref{prest}) as 
\begin{eqnarray}
\rho (r) &=& \frac{f(T_{0})}{16\pi} +\frac{f_{T}(T_{0})}{8\pi r^{2}}-\frac{f_{T}(T_{0})T_{0}}{8\pi}\label{dens2}\; ,\\
p_{r} (r) &=& -\rho (r)\label{presr2}\; ,\\
p_{t} (r) &=& \frac{f_{T}(T_{0})T_{0}}{8\pi}-\frac{f(T_{0})}{16\pi}\label{prest2}\; .
\end{eqnarray} 
In fact, we can show that this is the first solution for a wormhole in the $f(T)$ theory. Wormholes solutions also appear in the theory $f(R)$ \cite{manuel6}.\par
Through (\ref{dens2}) and (\ref{presr2}), we see that the    matter content solution is singular in $r=0$. This solution seems to obey to a condition of dark energy, $p_{r}(r)=-\rho(r)$, but which is quite different, with $p_{t}$ constant and different from $p_r$, thus being anisotropic. In fact, we can show that this is the first solution of a wormhole for the $f(T)$ theory.\par

The conditions of existence of a wormhole are the following. First we define the metric (\ref{ele})  in terms of the proper length $l$ as
\begin{equation}
dS^{2}=e^{a(r)}dt^{2}-dl^{2}-r^{2}(l)d\Omega^{2}\label{elw}\;,
\end{equation}
where $a(r)$ is denoted redshift function, and through the redefinition $\beta (r)=r\left[1-e^{-b(r)}\right]$, with $b(r)$ being the metric function given in (\ref{ele}), $\beta(r)$ is called  shape function. Therefore, the conditions of existence of a traversable wormhole are \cite{visser}; a) the function $r(l)$ must possess a minimum value $r_{0}$ for $r$, which imposes ${d^2r}(l)/dl^{2}>0$; b) $\beta(r_{0})=r_{0}$;  c) $a(r_{0})$ has a finite value; and finally d) $\beta(r)/dr|_{r=r_{0}}\leqslant  1$.\par

For our solution (\ref{sol2}), making $dr/dl=\sqrt{e^{-b(r)}}=0$, we obtain the following minimum value  $r_{0}=\sqrt[3]{-6c_{0}/T_{0}}$. The redshift function $a(r)$ has a finite value in  $r_{0}$, and $\beta (r_{0})=r_{0}$. Imposing the condition  $d\beta (r)/dr|_{r=r_{0}}\leqslant 1$, one gets  $T_{0}\geqslant 0$, which implies that $c_{0}<0$ ($T_{0}>0$), for getting  $r_{0}>0$. Therefore, the solution (\ref{sol2}) is a traversable wormhole. This wormhole connects two non asymptotically flat regions, which is asymptotically anti-de Sitter (AdS) for $T_{0}>0$. The wormholes solutions asymptotically (A)dS had been developed in \cite{lemos}. The most interesting aspect in this solution is that the energy conditions are satisfied here. In General Relativity, wormholes solutions violate some energy conditions, but here, we get a solution that satisfies all energy conditions. We show this as follows:
From the equation (\ref{presr2}), one gets  $\rho(r)+p_{r}(r)=0$, which satisfies the condition  $\rho (r)+p_{r}(r)\geqslant 0$. From the equations (\ref{dens2}) and  (\ref{prest2}), we obtain $\rho (r)+p_{t}(r)=f_{T}(T_{0})/8\pi r^2$, which, for $f_{T}(T_{0})\geqslant 0$, satisfies the condition $\rho (r)+p_{t}(r)\geqslant 0$. Through the equation  (\ref{dens2}), for $f(T_{0})\geqslant 2T_{0}f_{T}(T_{0})$, always $\rho (r)\geqslant 0$, and for  $f(T_{0})< 2T_{0}f_{T}(T_{0})$, if  $r_{0}\leqslant r\leqslant r_{1}$, with $r_{1}=\sqrt{2f_{T}(T_{0})/\left[2T_{0}f_{T}(T_{0})-f(T_{0})\right]}$, we obtain  $\rho (r)\geqslant 0$. Thus, the weak energy condition (WEC) and the null energy condition (NEC) are satisfied.   
\par 
In both solutions (\ref{sol1}) and (\ref{sol2}) presented here, for the constant torsion scalar, the value of $T_{0}$ and the functions $f(T_{0})$ and $f_{T}(T_{0})$ cannot be arbitrary as suggested by  Boehmer \cite{boehmer}.

%%%%%%%%%%%%%%%%%%%%%%%%%%%%%%%%%%%%%

\item  The second case we present here is that for which the radial pressure (\ref{presr}) is taken to be identically null. This has been done originally in GR by Florides \cite{florides} and used  later by Boehmer et al \cite{boehmer1}. In this case the equation  (\ref{presr}) reads
\begin{eqnarray}
f(T)=2f_{T}(T)\left(T-\frac{1}{r^2}\right)\label{cond3}\; .
\end{eqnarray}
Assuming that the function  $f(T)$ is given for the imposition (\ref{impf1}), we obtain 
\begin{eqnarray}
T(r)=\frac{a_{0}}{a_{1}}+\frac{2}{r^2}\label{t1}\,.
\end{eqnarray}

We distinguish the following two interesting sub-cases:
%%%%%%%%%%%%%%%%%%%%%%%%%%%%%%
\begin{enumerate}

%%%%%%%%%%%%%%%%%%%%%%%%%%%%%%

\item For the case in which $a(r)$ obeys  the condition (\ref{cond1}), we calculate $b(r)$ from the expression of the torsion scalar (\ref{te}), yielding the line element
\begin{equation}
dS^{2}=\frac{r_{0}}{r}e^{c_{0}r}dt^{2}-\left(\frac{a_{0}r}{2a_{1}c_{0}}+\frac{1}{c_{0}r}\right)^{-1}dr^2-r^2d\Omega^2\; .\label{sol3}
\end{equation} 
The energy density, radial and tangential pressures are given by
\begin{eqnarray}
\rho (r) &=& \frac{a_0}{16\pi} +\frac{a_1}{8\pi r^{2}}-\frac{a_0}{8\pi c_{0}r}\label{dens3}\; ,\\
p_{r} (r) &=& 0\label{presr3}\; ,\\
p_{t} (r) &=& -\frac{3a_{0}}{64\pi}+\frac{a_{0}c_{0}r}{64\pi}+\frac{1}{32\pi r}\left(\frac{a_{0}}{c_{0}}+a_{1}c_{0}\right)-\frac{a_{1}}{32\pi r^2}\label{prest3}\; .
\end{eqnarray} 
This solution is clearly singular in $r=0$. This solution is a traversable wormhole that connects two non asymptotically flat regions. Comparing (\ref{ele}), (\ref{elw}) and (\ref{sol3}), we observe that for $dr/dl=0 $, the radial coordinate $r$ possesses a minimum value  $r_{1}=\sqrt{-2a_{1}/a_{0}}$, which implies that  $sign(a_{1}/a_{0})=-1$. The redshift function $a(r)$ has a finite value in $r_{1}$, and the shape function obeys  $\beta (r_{1})=r_{1}$. Imposing the condition $d\beta (r)/dr|_{r=r_{1}}\leqslant 1$, and as $r_{1}>0$, one gets $c_{0}< 0$. As $p_{r}(r)=0$, the energy conditions  $\rho (r)\geqslant 0$ and $\rho (r)+p_{r}(r)\geqslant 0$ are the same, ie, are equivalent to  $\rho (r)\geqslant 0$. Taking into account the both conditions $\rho (r)\geqslant 0$ and $\rho (r)+p_{t}(r)\geqslant 0$, we get two possibilities to satisfy these energy conditions: for $a_{0}>0$ ($a_{1}<0$), we must get $3/|c_{0}|\leqslant r_{1}\leqslant r\leqslant r_{2}$, with  $r_{2}=(1/|c_{0}|)\left(\sqrt{1+2|a_{1}/a_{0}|c_{0}^2}-1\right)$; for $a_{0}<0$ ($a_{1}>0$), we must obtain  $3/|c_{0}|\leqslant r_{1}\leqslant r<+\infty$.    

%%%%%%%%%%%%%%%%%%%%%%%%%%%%%%%%

\item Considering now the condition (\ref{cond2}), we obtain from  (\ref{te}),
\begin{equation}
\left(e^{-b}\right)^{\prime}+\left(\frac{1}{r}\right)\left(e^{-b}\right)-\frac{r}{2}T(r)=0\label{eq2}\; ,
\end{equation}
which, using (\ref{t1}), leads to the line element
\begin{equation}\label{sol4}
dS^{2}=\left(1+\frac{c_{0}}{r}+\frac{a_{0}}{6a_{1}}r^2\right)dt^2-\left(1+\frac{c_{0}}{r}+\frac{a_{0}}{6a_{1}}r^2\right)^{-1}dr^2-r^2d\Omega^2\; .
\end{equation}
This is similar to  S-(Anti)dS solution when we fix $c_{0}=-2M$ and $\frac{a_0}{2a_1}=-\Lambda$. Note that, once again, Wang \cite{wang} found a similar solution in which the difference is about the fixation of the function $f(T)=T=-2\Lambda$, which is not the case here, because $T(r)$ in (\ref{t1}) depends on the radial coordinate and is not constant. The density, the radial and tangential pressures are identically zero
\begin{equation}
\rho (r)=p_{r}(r)=p_{t}(r)=0\label{dens4}\,.
\end{equation}
This interior and regular solution is identified as the vacuum of matter, ie, $\mathcal{T}^{\nu}_{\mu}=0$ in (\ref{em}). But as mentioned in \cite{ferraro2}, the frame in (\ref{ele}) represents a material vacuum but with a non vanishing torsion scalar, given in (\ref{t1}). This is an aspect of the $f(T)$ theory, which is no longer invariant under local Lorentz transformations.
%%%%%%%%%%%%%%%%%%%%%%%%%%%%%%%%%%%%%%%%%%%%%%%%

\item Another way to regain these results of the previous sub-item is differentiating the equation (\ref{cond3}) with respect to $T$, and considering (\ref{impf1}), which results into
\begin{eqnarray}
\frac{dr^{-2}}{dT}=-\frac{1}{2}\label{eq3}\;.
\end{eqnarray} 
Integrating (\ref{eq3}) in the limits of $r_{0}$ to $r$ and  $T_{0}$ to $T$, we get
\begin{eqnarray}
T(r)=T_{0}-\frac{2}{r_{0}^2}+\frac{2}{r^2}\label{t2}\;,
\end{eqnarray}
in which, choosing  $T_{0}=(2/r_{0}^2)+(a_{0}/a_{1})$, the two previous cases are recovered.
%%%%%%%%%%%%%%%%%%%%%%%%%%%%%%%%%%%
\end{enumerate}
%%%%%%%%%%%%%%%%%%%%%%

\item Taking the case in which the radial pressure is a constant,
\begin{equation}
p_{r}(r)=p_{r}\,\,\,,\label{cond4}
\end{equation}
where $p_{r}\in \mathcal{R}$, with the imposition  (\ref{impf1}), we obtain the torsion scalar on the form 
\begin{equation}
T(r)=\frac{16\pi p_{r}+a_{0}}{a_{1}}+\frac{2}{r^2}\label{t2}\;.
\end{equation}

Then, we have two possibilities 
%%%%%%%%%%%%%%%%%%%%%%%
\begin{enumerate}
%%%%%%%%%%%%%%

\item When we use the condition (\ref{cond1}), the line element is given by 
\begin{eqnarray}
dS^2=\frac{r_{0}}{r}e^{c_{0}r}dt^2-\left[\frac{(16\pi p_{r}+a_{0})}{2a_{1}c_{0}}r+\frac{1}{c_{0}r}\right]^{-1}dr^2-r^2d\Omega^2\label{sol5}\;.
\end{eqnarray}

The  energy density,  the radial and tangential pressures are given by 
\begin{eqnarray}
\rho (r) &=& \frac{a_0}{16\pi} +\frac{a_1}{8\pi r^{2}}-\frac{a_0}{8\pi c_{0}r}-\frac{2p_{r}}{c_{0}r}\label{dens5}\; ,\\
p_{r} (r) &=& p_{r}\label{presr5}\; ,\\
p_{t} (r) &=& -\frac{3a_{0}}{64\pi}+\frac{p_{r}}{4}+\frac{a_{0}c_{0}r}{64\pi}+\frac{1}{32\pi r}\left(\frac{a_{0}}{c_{0}}+a_{1}c_{0}\right)-\frac{a_{1}}{32\pi r^2}+\frac{p_{r}}{2c_{0}r}+\frac{c_{0}p_{r}}{4}r\label{prest5}\; .
\end{eqnarray}
This solution is singular in $r=0$, and generalizes the solution (\ref{sol3}). When $p_{r}=0$ in  (\ref{sol5})-(\ref{prest5}), we regain (\ref{sol3})-(\ref{prest3}). This solution is a traversable wormhole that connects two non asymptotically flat regions. The demonstration follows the same steps as in (\ref{sol3}), substituting $a_0$ by $a_{0}\rightarrow a_{0}+16\pi p_{r}$. 
%%%%%%%%%%%%%%%%%%%%%%%%%%%%%

\item For the condition (\ref{cond2}), we have the equation (\ref{eq2}), which integrated, yields the line element 
\begin{eqnarray}
dS^2=\left[1+\frac{c_{0}}{r}+\frac{(16\pi p_{r}+a_{0})}{6a_{1}}r^2\right]dt^2-\left[1+\frac{c_{0}}{r}+\frac{(16\pi p_{r}+a_{0})}{6a_{1}}r^2\right]^{-1}dr^2-r^2d\Omega^2\label{sol6}\;.
\end{eqnarray}

The density, the radial and tangential pressures are given by 
\begin{eqnarray}
p_{r} (r)=-\rho (r)=p_{t} (r)=p_{r} \label{dens6}\; .
\end{eqnarray} 
 This interior solution is regular for all values of the radial coordinate $r$. It is similar to the S-(A)dS's solution  when $c_{0}=-2M$ and $(16\pi p_{r}+a_{0})/2a_{1}=-\Lambda$. It also obeys some dark energy condition between the density, the radial and tangential pressures in (\ref{dens6}). Our model starts considering an anisotropic spacetime with a constant radial pressure. Due to the choice of the quasi-global coordinate condition (\ref{cond2}), the equations lead to the isotropization of the spacetime, where $p_{r}=p_{t}$. This solution yields (\ref{sol4})-(\ref{dens4}) for $p_{r}=0$ in  (\ref{sol6})-(\ref{dens6}). 
 For the case in which the solution looks like S-dS (or S-AdS), we will obtain, according to equation (\ref{t2}), a possibility of a change in the sign of the torsion scalar. The radial pressure (equal to the tangential one) should always be a negative constant for obtaining the positivity of the energy density $\rho$ in (\ref{dens6}).
%%%%%%%%%%%%%%%%%%%%%
\end{enumerate}
%%%%%%%%%%%%%%%%%%%%%%

\item A condition taken directly from the equations of motion (\ref{presr}), is when the radial pressure is proportional to the algebraic function $f(T)$. For the particular case of constant radial pressure, we regain the imposition (\ref{impt1}) or (\ref{imptf1}), treated in the previous subsection. Then, we have now to treat the case where $f(T)$ obeys the imposition (\ref{impf1}), for which, assuming the dependence of the radial pressure as
\begin{equation}
p_{r}(r)=\frac{f(T)}{c_{0}}\label{cond5}
\end{equation}
and using the equation (\ref{presr}), we get 
\begin{equation}
T(r)=\frac{a_{0}}{a_{1}}\left(\frac{c_{0}+16\pi}{c_{0}-16\pi}\right)+\frac{2c_{0}}{(c_{0}-16\pi)r^2}\;.\label{t3}
\end{equation}
%%%%%%%%%%%%%%%%%%%%%%%%%%%%%%%%%%%%%%%%%%%%%%%%

By considering the imposition (\ref{impf1}) and integrating within the limits of $r_{0}$ to $r$ and $T_{0}$ to $T$, with $T_{0}=(a_{0}/a_{1})[(c_{0}+16\pi)/(c_{0}-16\pi)]+[2c_{0}/(c_{0}-16\pi)r_{0}^2]$,  the result (\ref{t3}) can be obtained using the condition (\ref{cond5}) in the equation of motion (\ref{presr}), and differentiating  with respect to the torsion scalar $T$. We then have two possibilities
%%%%%%%%%%%%%%%%
\begin{enumerate}
%%%%%%%%%%

\item For the condition (\ref{cond1}), the solution is given by 
\begin{eqnarray}
dS^2 &=& \frac{r_{0}}{r}e^{c_{1}r}dt^2-\left[\left(\frac{c_{0}+16\pi}{c_{0}-16\pi}\right)\frac{a_{0}}{2a_{1}c_{1}}r+\frac{c_{0}}{c_{1}(c_{0}-16\pi)r}\right]^{-1}dr^2-r^2d\Omega^2\label{sol7}\;,\\
\rho (r) &=& \frac{a_{0}}{(16\pi-c_{0})}\left(1-\frac{c_{0}}{16\pi}+\frac{1}{c_{1}r}\left[2+\frac{c_{0}}{8\pi c_{1}}\right]\right)+\frac{a_{1}}{(16\pi-c_{0})r^2}\left(2- \frac{c_{0}}{8\pi}\right)\label{dens7}\;,\\
p_{r}(r) &=&\frac{2}{(c_{0}-16\pi)}\left( a_{0}+\frac{a_{1}}{r^2}\right) \label{presr7}\,,\\
p_{t}(r) &=& \frac{a_{0}}{2(16\pi-c_{0})}\left[\frac{3c_{0}}{32\pi}-\frac{5}{2}-\frac{1}{c_{1}r}-\frac{c_{0}}{4\pi c_{1} r}-\frac{c_{1}r}{2}\left(1+\frac{c_{0}}{16\pi}\right)\right]\nonumber\\
&&\frac{a_{1}c_{0}}{32\pi (c_{0}-16\pi)r}\left(\frac{1}{r}-c_{1}\right)\label{prest7}\; .
\end{eqnarray}

This solution is singular in $r=0$. With the limit  $c_{0}\rightarrow +\infty$ in (\ref{sol7})-(\ref{prest7}), we regain the solution  (\ref{sol3})-(\ref{prest3}), where $p_{r}=0$. This is a traversable wormhole that connects two non asymptotically flat regions. The demonstration follows the same steps as in (\ref{sol3}).

%%%%%%%%

\item For the condition (\ref{cond2}), the solution is given by 
\begin{eqnarray}
dS^2 &=& \left[\frac{c_{0}}{c_{0}-16\pi}+\frac{c_{1}}{r}+\frac{a_{0}}{6a_{1}}\frac{(c_{0}+16\pi)}{(c_{0}-16\pi)}r^2\right]dt^2\nonumber\\
&&-\left[\frac{c_{0}}{c_{0}-16\pi}+\frac{c_{1}}{r}+\frac{a_{0}}{6a_{1}}\frac{(c_{0}+16\pi)}{(c_{0}-16\pi)}r^2\right]^{-1}dr^2-r^2d\Omega^2\label{sol8}\;,\\
\rho (r) &=& -\frac{2a_{0}}{c_{0}-16\pi}-\frac{2a_{1}}{(c_{0}-16\pi)r^2}\label{dens8}\;,\\
p_{r}(r) &=& -\rho (r)\label{presr8}\,,\\
p_{t}(r) &=& \frac{2a_{0}}{c_{0}-16\pi}\label{prest8}\; .
\end{eqnarray}
This is a solution similar to that of S-(A)dS. The constants $c_{0}, a_{0}$ and $a_{1}$ should take values such that the positivity of the energy density is guaranteed in (\ref{dens8}). This is another singular solution at $r=0$. Again, for the limit $ c_{0}\rightarrow + \infty $ in (\ref{sol8})-(\ref{prest8}), we regain the solution (\ref{sol6})-(\ref{dens6}), where $p_{r}=0$.

%%%%%%%%%%%
\end{enumerate}
%%%%%%%%%%%%%%%%%%

\item Making use of the condition 
\begin{equation}
p_{r}(r)=\frac{\eta}{8\pi r^2}f_{T}(T)\label{cond6}\;,
\end{equation}
where $\eta\in\mathcal{R}$,  the equation  (\ref{presr}) leads to 
\begin{eqnarray}
T(r)=\frac{a_{0}}{a_{1}}+\frac{2(1+\eta)}{r^2}\label{t4}\;.
\end{eqnarray}
This result can be obtained by replacing the condition (\ref{cond6}) in the equation of motion (\ref{presr}), dif\-ferentiating with respect to the torsion scalar $T$, imposing (\ref{impf1}) and integrating within the limits $r_{0}$ to $r$ and  $T_{0}$ to  $T$, with $T_{0}=(a_{0}/a_{1})+[2(1+\eta)/r_{0}^2]$. Then, we have the following possibilities
%%%%%%%%%%%%%
\begin{enumerate}
%%%%%%%%%%%%

\item  For the condition (\ref{cond1}), we get the following solution 
\begin{eqnarray}
dS^2 &=& \frac{r_{0}}{r}e^{c_{0}r}dt^2-\left[\frac{a_{0}}{2a_{1}c_{0}}r+\frac{1+\eta}{c_{0}r}\right]^{-1}dr^2-r^2d\Omega^2\label{sol9}\;,\\
\rho (r) &=& \frac{a_{0}}{8\pi}\left(\frac{1}{2}-\frac{1}{c_{0}r}\right)+\frac{a_{1}}{8\pi r^2}\label{dens9}\;,\\
p_{r}(r) &=& \frac{\eta a_{1}}{8\pi r^2}\label{presr9}\,,\\
p_{t}(r) &=& \frac{a_{0}}{32\pi}\left(-\frac{3}{2}+\frac{1}{c_{0}r}+\frac{c_{0}r}{2}\right)+\frac{a_{1}(1+\eta)}{32\pi r}\left(c_{0}-\frac{1}{r}\right)\label{prest9}\; .
\end{eqnarray}

This is a singular solution in $r=0$. With $\eta =0$ in (\ref{sol9})-(\ref{prest9}), we re-obtain  (\ref{sol3})-(\ref{prest3}), where $p_{r}=0$. This is a new traversable wormhole solution that connects two non asymptotically flat regions.  This can be easily observed following the same steps as in (\ref{sol3}). 

%%%%%%%%%%%%%%

\item  For the condition (\ref{cond2}), we have the following solution 
\begin{eqnarray}
dS^2 &=& \left[(1+\eta)+\frac{c_{0}}{r}+\frac{a_{0}}{6a_{1}}r^2\right]dt^2-\left[(1+\eta)+\frac{c_{0}}{r}+\frac{a_{0}}{6a_{1}}r^2\right]^{-1}dr^2-r^2d\Omega^2\label{sol10}\;,\\
\rho (r) &=&-p_{r}(r)= -\frac{\eta a_{1}}{8\pi r^2}\;,\; p_{t}(r) =0 \label{dens10}\;.
\end{eqnarray}
This solution resembles to that of S-(A)dS. Once again, we have a singular solution at $r= 0$. When we make use of $\eta = 0$ in (\ref{sol10})-(\ref{dens10}), we regain the solution (\ref{sol4})-(\ref{dens4}) for $p_{r} = 0$. The energy density (\ref{dens10}) obeys the restriction $\eta a_{1} <0 $, for ensuring  its  positivity. But the relation $p_{r}(r)=-\rho(r)$ in this case does not provide us a model like dark energy, due to $ p_{t}\neq p_{r}$, then being anisotropic. 

%%%%%%%%%
\end{enumerate}
%%%%%%%%%%
\end{enumerate}

%%%%%%%%%%%%%%%%%%%%%%%%%%%%%%%%%%%%%%%%%%%%%%%%%%%%%%%%%%%%%%%%%%%%%%%%%%%%%%%%%%%%%%%%%%%%%%%%%%

%%%%%%%%%%%%%%%%%%%%%%%%%%%%%%%%%%%%%%%%%%%%%%%%%%%%%%%%%%%%%%%%%%%%%%%%%%%%%%%%%%%%%%%%%%%%%%
\section{\large Conclusion}

%%%%%%%%%%%%%%%%%%%%%%%%%%%%%%%%%%%%%%%%%%%%%%%%%%%%%%%%%%%%%%%%%%%%%%%%%%%%%%%%%%%%%%%%%%%%%%%%%%%%%%%%%%%%%%%%%%%%%%%%
The equations of motion for the $f(T)$ theory are presented in section $2$, showing a surprising result. The fact is that there is an off-diagonal equation for a static case with spherical symmetry. This was first shown by Boehmer et al \cite{boehmer}, and now confirming the conclusion made ​in \cite{yapiskan} for a diagonal static metric. But Boehmer et al still make another statement that, in the case of non diagonal tetrad $e^{a}_{\mu}$, the constraint equation (\ref{impos}) no longer holds. Thus, the algebraic function $f(T)$ and its derivative are arbitrary, only with the constraint of positivity on the energy density in (\ref{dens}).\par

Through the consideration that the torsion of the Weitzenbock's geometry is constant, we obtained new spherically symmetric static solutions for the $f(T)$ theory, fixing a generalization of the case of Boehmer et al \cite{boehmer}, in (\ref{cond1}), and a new assumption for which $a^{\prime}(r)=-b^{\prime}(r)$. The solution which generalizes the case of Boehmer et al presents a signature such that the coordinates $t$ and $r$ are time-like and space-like respectively. There is other solution similar to the S-(Anti)dS's one, obtained by Wang \cite{wang}, for which the unique difference is fixing the function $f(T)=T=-2\Lambda$. Note that here, this assumption does not make free $f(T_0)$. In fact, this solution is a traversable wormhole which fixes the positivity of the torsion scalar and the algebraic function $f_{T}(T_{0})$.\par
We also constructed four conditions that fix the matter content as a function of $r$. The first is when the radial pressure is identically zero, the second when it is a real constant, the third is when it is proportional to the algebraic function $f(T)$, and the last is when it is proportional to $f_{T}(T)$ multiplied by $\eta/8\pi r^2 $. Through the  coordinates conditions (\ref{cond1}) and (\ref{cond2}), we obtain new classes of static black holes solutions and wormholes to the $f(T)$ theory.\par 
Using the simplification methods for the equations of motion in GR \cite{florides,jamil,prisco}, we shown that it is possible to obtain new spherically symmetric static anisotropic solutions for $f(T)$ theory. With this, we hope that others conditions and symmetries reveal new solutions, even analogous to GR, as we see here, or also shown in \cite{stephane}.\par
An important point to be observed in this work is that, according to the equation (\ref{dens}),  the torsion $T$, the functions $f(T)$ and $f_{T}$ cannot be arbitrary functions because conditions are required on these functions following the energy conditions in $f(T)$ gravity. We presented here a brief discussion about some particular  wormholes solutions. However, this discussion could be made in more detail. We propose to present this aspect in a future work.

%%%%%%%%%%%%%%%%%%%%%%%%%%%%%%%%%%%%%%%%%%%%%%%%%%%%%%%%%%%%%%%%%%%%%%%%%%%%%%%%%%%%%%%%%%%%%%%%%%%%%%%%%%%%%%%%%%%%%%%%%%%%%%%%%
\vspace{1cm}

{\bf Acknowledgement:}   M. H. Daouda thanks CNPq/TWAS for financial support. M. E. Rodrigues  thanks  UFES for the hospitality during the development of this work. M. J. S. Houndjo thanks  CNPq for partial financial support.

%%%%%%%%%%%%%%%%%%%%%%%%%%%%%%

%%%%%%%%%%%%%%%%%%%%%%%%%%%%%%%%%%%%%%%%%%%%%%%%%%%%%%%%%%%%%%%%%%%%%%%%%%%%%%%%%%%%%%%%%%%%%%
\end{document}